\documentstyle[aps,prl,preprint,floats,epsfig]{revtex}

\begin{document}

\preprint{\tighten\vbox{
                        \hbox{\hfil CLEO CONF 01-1}
}}

\title{Mixing and CP Violation in the Decay of \\
Neutral D Mesons at CLEO.}  

\author{CLEO Collaboration}
\date{February 5, 2001}

\maketitle
\tighten

\begin{abstract} 
We present preliminary results of several analyses 
searching for the effects of {\em CP} violation and mixing 
in the decay of $D^0$ mesons.  
We find no evidence of {\em CP} asymmetry in five different 
two-body decay modes of the $D^0$ to pairs of light pseudo-scalar mesons:
$A_{CP}(K^+ K^-) = (+0.05 \pm 2.18 \pm 0.84)\% $,
$A_{CP}(\pi^+ \pi^-) = (+2.0 \pm 3.2 \pm 0.8)\% $,
$A_{CP}(K^0_{\rm S} \pi^0) = (+0.1 \pm 1.3)\%$, 
$A_{CP}(\pi^0 \pi^0) = (+0.1 \pm 4.8)\%$ and 
$A_{CP}(K^0_{\rm S} K^0_{\rm S}) = (-23 \pm 19)\%$.
We present the first measurement of the rate of wrong-sign 
$D^0 \rightarrow K^+ \pi^- \pi^0$ decay:
$R_{WS} = 0.0043^{+0.0011}_{-0.0010} \pm 0.0007$.
Finally, we describe a measurement of the mixing parameter $y_{CP}=
{\Delta\Gamma\over 2 \Gamma}$ 
by searching for a lifetime difference between the {\em CP} 
neutral $K^+ \pi^-$ final state and the 
{\em CP} even $K^+K^-$ and $\pi^+\pi^-$ final states.  
Under the assumption that {\em CP} is conserved we find 
$y_{CP} = -0.011 \pm 0.025 \pm 0.014$.

\end{abstract}
\newpage

{
\renewcommand{\thefootnote}{\fnsymbol{footnote}}


\begin{center}
D.~Cronin-Hennessy,$^{1}$ A.L.~Lyon,$^{1}$ E.~H.~Thorndike,$^{1}$
T.~E.~Coan,$^{2}$ V.~Fadeyev,$^{2}$ Y.~S.~Gao,$^{2}$
Y.~Maravin,$^{2}$ I.~Narsky,$^{2}$ R.~Stroynowski,$^{2}$
J.~Ye,$^{2}$
% ...
\end{center}
 
\small
\begin{center}
$^{1}${University of Rochester, Rochester, New York 14627}\\
$^{2}${Southern Methodist University, Dallas, Texas 75275}\\
% ...
\end{center}
 
\setcounter{footnote}{0}
}
\newpage

\section{Introduction and Motivation}\label{sec:intro}
The study of mixing
in the $K^0$ and $B_d^0$ sectors has provided a wealth of information
to guide the form and content of the Standard Model.  In the framework of the 
Standard Model, mixing in the charm meson sector is predicted to 
be small~\cite{harry}, making this an excellent place to
search for non-Standard Model effects.  Similarly, measurable
{\em CP} violation (CPV) phenomena in strange~\cite{KTeV,NA48} and 
beauty~\cite{BaBar,Belle,CDF} mesons are the impetus for many current and future 
experiments~\cite{AGS,KAMI,BTeV,LHCb}.
The Standard Model predictions for CPV for charm mesons 
are of the order of $0.1\%$~\cite{Bucella}, with one recent conjecture of nearly 
$1\%$~\cite{Bigi}.  Observation of CPV in charm mesons exceeding the percent
level would be strong evidence for non-Standard Model processes.

A $D^0$ can evolve into a $\overline{D^0}$ through ``ordinary'' on-shell 
intermediate states, or through off-shell intermediate states such as as
those that might be present due to new physics.  We denote the
amplitude through the former (latter) states by $-iy$ $(x)$, in
units of $\Gamma_{D^0} / 2$~\cite{Mixing}.  The Standard Model contributions to
$x$ are suppressed to $|x| \approx \tan^2 \theta_C \approx 5\%$ and the
Glashow-Illiopolous-Maiani~\cite{GIM} cancelation could further
suppress $|x|$ down to $10^{-6} - 10^{-2}$.  Many non-Standard Model processes
could lead to $|x| > 1\%$.  Contributions to $x$ at this level could result
from the presence of new particles with masses as high as 100 --
1000 TeV~\cite{Leurer}.  Signatures of new physics include $|x| \gg |y|$ and
{\em CP} violating interference between $x$ and $y$ or between $x$ and a 
direct decay amplitude.

Wrong sign (WS) processes, such as $D^0 \rightarrow K^+ \pi^- \pi^0$, can proceed
directly through doubly Cabbibo-suppressed decay (DCSD) or through
mixing and subsequent Cabbibo favored decay (CFD).  Both DCSD and
mixing followed by CFD can contribute to the time integrated WS
rate, $R_{\rm WS} = (r + \bar{r})/2$ and the inclusive {\em CP} asymmetry 
$A = (r - \bar{r})/(r + \bar{r})$, where $r = \Gamma \left( D^0 \rightarrow 
f \right) /\Gamma \left( \overline{D^0} \rightarrow 
f \right)$, $\bar{r}$ is the charge conjugated quantity, and $f$ is a 
wrong-sign final state, such as $K^+ \pi^- \pi^0$.

The different contributions to $R_{\rm WS}$ and $A$ can be separated by 
studying the proper decay time dependence of WS final states, 
as we have done in $D^0 \rightarrow K^+ \pi^-$~\cite{kpi}, and has been done by
FOCUS in \cite{FOCUS}.  The differential WS rate
relative to the right-sign (RS) process, in time units of the
mean $D^0$ lifetime, $t_{D^0} = (415 \pm 4) {\rm fs}$~\cite{PDG}, is 
$r(t) \equiv [R_{\rm D} + \sqrt{R_{\rm D}} y^\prime t + 1/4 ({x^\prime}^2 + 
{y^\prime}^2 ) t^2 ] e^{-t}$~\cite{Treiman},
where $y^\prime \equiv y \cos \delta - x \sin \delta$ and $x^\prime \equiv x 
\cos \delta
+ y \sin \delta$, where $R_{\rm D}$ is the relative rate of DCSD and
$\delta$ is the strong phase between the DCSD and CFD
amplitudes.  There are theoretical arguments that $\delta$ should be 
small~\cite{Wolf}, 
although this should not be taken for granted.  The coefficient of the term
quadratic in $t$ is proportional to the relative rate of mixing,
$R_{\rm M} \equiv 1/2 (x^2 + y^2) = 1/2 ({x^\prime}^2 + {y^\prime}^2 )$.


\section{General Experimental Method}\label{genmethod}

All of the analyses discussed herein, unless otherwise stated,
use the same data set and reconstruction techniques described
below.  The data set was accumulated between February 1996 and
February 1999 and corresponds to $9.0\  {\rm fb}^{-1}$ of $e^+ e^-$ collision
data at $\sqrt{s} \approx 10.6$ GeV provided by the Cornell Electron
Storage Ring (CESR).  The data were recorded by the CLEO II
detector~\cite{cleoii} upgraded with the installation of a silicon vertex
detector (SVX)~\cite{svx} and by changing the drift chamber gas from
an argon-ethane mixture to a helium-propane mixture~\cite{dr}.  The 
upgraded configuration is referred to as CLEO II.V.      
        
The Monte Carlo simulation of the CLEO II.V detector was based upon
GEANT~\cite{GEANT}, and simulated events were processed in the same manner
as the data.

The $D^0$ candidates are reconstructed through the decay sequence
$D^{\star +} \rightarrow D^0 \pi^+_{\rm s}$~\cite{charge}.  
The charge of the slow pion ($\pi^+_{\rm s}$) tags the
flavor of the $D^0$ candidate at production.  The charged daughters
of the $D^0$ are required to leave hits in the SVX and these tracks
are constrained to come from a common vertex in three dimensions.
The trajectory of the $D^0$ is projected back to its
intersection with the CESR luminous region to obtain the $D^0$
production point.  The $\pi^+_{\rm s}$ is refit with the requirement that
it come from the $D^0$ production point, and the confidence level of the $\chi^2$ 
of this refit is used to reject background.  

The energy release in the $D^\star \rightarrow D^0 \pi^+_{\rm s}$ decay, 
$Q \equiv M^\star - M - m_\pi$, 
obtained from the above technique is observed to have a width of 
$\sigma_Q = 190 \pm 2$ keV~\cite{qwidth}, which is a combination of
the intrinsic width and our resolution,
where $M$ and $M^\star$ are the reconstructed 
masses of the $D^0$ and $D^{\star +}$ candidates respectively, and
$m_\pi$ is the charged pion mass.  The reconstruction technique 
discussed above has also been used by CLEO to measure the $D^{*+}$
intrinsic width, $\Gamma_{D^{*+}} = 96\pm 4\pm 22$ keV 
(preliminary)~\cite{GammaD*}.  
In the mixing analyses described below, the distribution of
candidates in the $Q$ vs $M$ plane are fit to determine both right-sign 
and wrong-sign yields.

We calculate $t$ using only the vertical component of the $D^0$
candidate flight distance.  This is effective
because the vertical extent of the CESR luminous region has
$\sigma_{\rm vertical} = 7 \mu$m~\cite{cinabro}.  
The resolution on the $D^0$ decay point
($x_v$, $y_v$, $z_v$) is typically $40 \mu$m in each dimension.  We measure 
the centroid of the CESR luminous region ($x_b$, $y_b$, $z_b$) using
$e^+ e^- \rightarrow q \bar{q}$ ($q=udscb$) events from sets of data 
with integrated luminosity of several pb$^{-1}$, obtaining a resolution on the 
centroid of less than $5 \mu$m.  We express $t$ as 
$t = M/p_{\rm vertical} \times (y_v - y_b)/(c \tau_{D^0})$, where 
$p_{\rm vertical}$ is
the vertical component of the total momentum of the $D^0$ candidate.
The error in $t$, $\sigma_t$, is typically $0.4$ (in $D^0$ lifetimes), 
although when the $D^0$ direction is near the horizontal plane
$\sigma_t$ can be large.

\section{{\em CP} violation in $D^0$ decay}\label{sec:kkpp}
{\em CP} Violation in charm meson decay is expected to be small in the
Standard Model, which makes charm meson decay a good place to look for
non-Standard Model effects.
Cabbibo suppressed charm meson decays have all the necessary ingredients
for {\em CP} violation -- multiple paths to the same final state and a
weak phase.  However, in order to get sizable {\em CP} violation, the 
final state interactions need to contribute non-trivial phase shifts
between the amplitudes.  Large final state interactions are a likely reason why
the prediction for the ratio of branching ratios of 
$(D^0 \rightarrow K^+ K^-) / (D^0 \rightarrow
\pi^+ \pi^-)$ yields a value roughly half of the observed
value~\cite{PDG}, hence these may provide a good hunting ground 
for {\em CP} violation.

Previous searches for mixing-induced~\cite{kpi} or direct~\cite{kspi0,PDG} 
{\em CP} violation in the neutral charm meson system have set
limits of $\sim 30\%$ or a few percent, respectively.  We present results of
searches for direct {\em CP} violation in neutral charm meson decay to pairs of
light pseudo-scalar mesons: $K^+ K^-$, $\pi^+ \pi^-$, $K^0_{\rm S} \pi^0$, $\pi^0
\pi^0$ and $K^0_{\rm S} K^0_{\rm S}$.

\subsection{Search for {\em CP} violation in $D^0 \rightarrow K^+ K^-$
and $D^0 \rightarrow \pi^+ \pi^-$ decay}

The asymmetry we want to measure,
\begin{eqnarray*}
A = \frac{\Gamma \left( D^0 \rightarrow f \right) - 
\Gamma \left( \overline{D^0} \rightarrow f \right)}
{\Gamma \left( D^0 \rightarrow f \right) +
\Gamma \left( \overline{D^0} \rightarrow f \right)}
\end{eqnarray*}
can be obtained from the asymmetry
\begin{eqnarray*}
A^f = \frac{\Gamma \left( D^{\star +} \rightarrow \pi^+_{\rm s} f \right) - 
\Gamma \left( D^{\star -} \rightarrow \pi^-_{\rm s} f \right)}
{\Gamma \left( D^{\star +} \rightarrow \pi^+_{\rm s} f \right) +
\Gamma \left( D^{\star -} \rightarrow \pi^-_{\rm s} f \right)}
\end{eqnarray*}
The slow pion and $D^0$ are produced by the {\em CP}-conserving strong 
decay of the $D^{\star +}$, so the slow pion serves as an unbiased flavor tag
of the $D^0$.  The decay asymmetry can be obtained from the apparent production
asymmetry shown above because the production of $D^{\star \pm}$ is 
{\em CP}-conserving.

\begin{figure}
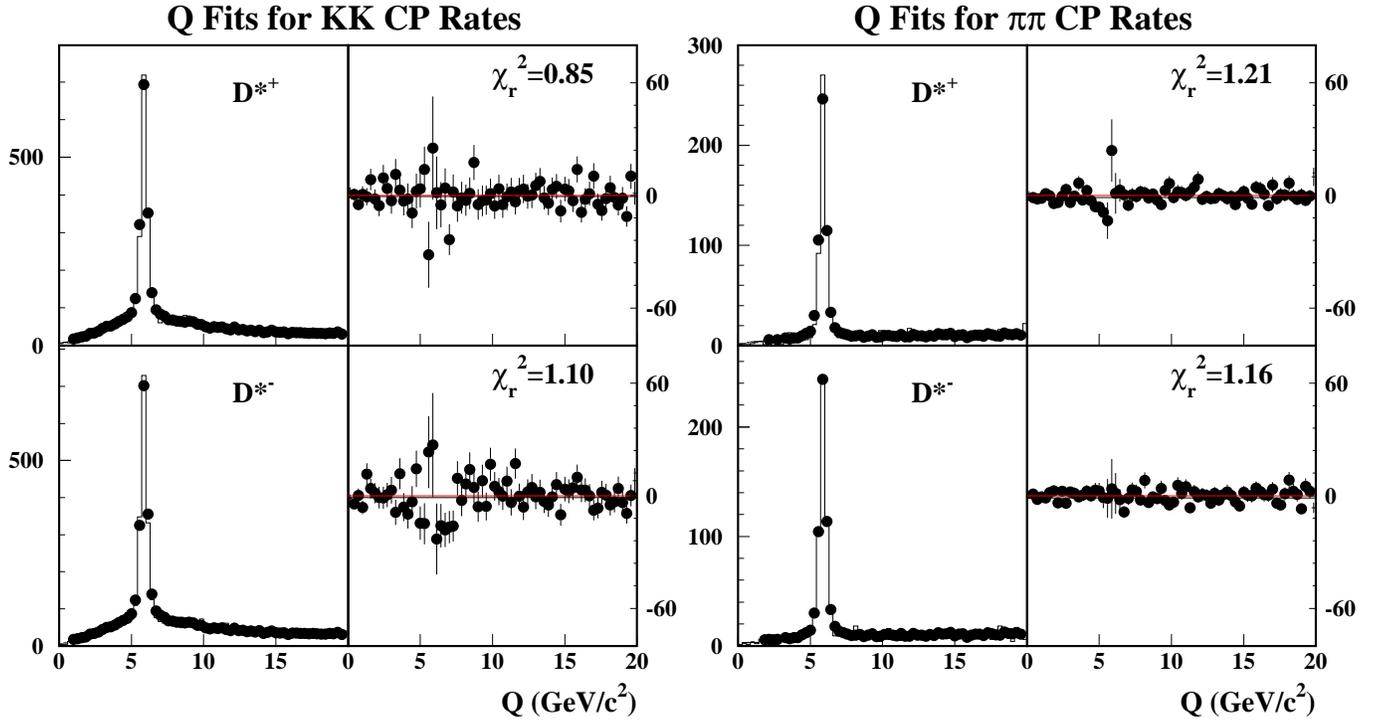

\centerline{
 \psfig{file=kk_cp_nomfit.epsi,width=3.5in}
 \psfig{file=pp_cp_nomfit.epsi,width=3.5in}
}
\vspace{0.5cm}
\caption{\label{fig:hill} $Q$ distributions for $D^0 \rightarrow K^+ K^-$ 
(left) and 
$D^0 \rightarrow \pi^+ \pi^-$ (right) candidates separated into samples tagged 
as $D^{*+}$ (top) and $D^{*-}$ (bottom).  Shown immediately to the right of 
each $Q$ distribution are the point by point residuals of the fit.
In each case the data points are shown with error bars and the solid 
line represents our fitting results.}
\end{figure}

The asymmetry result is obtained by fitting the energy release ($Q$) spectrum
of the $D^{\star +} \rightarrow D^0 \pi^+_{\rm s}$ events.  The $D^0$ mass
spectra are fit as a check.  The background-subtracted $Q$ spectrum is fit 
with a signal shape obtained from $K^+ \pi^-$ 
data and a background shape determined 
using Monte Carlo.  Figure ~\ref{fig:hill} shows our fits to both 
$D \rightarrow \pi^+ \pi^-$ and $D \rightarrow K^+ K^-$ distributions
found in data, for both signs of the slow pion flavor tag.
The parameters of the slow pion dominate the $Q$ distribution, 
so all modes have the same shape.
We do the fits in bins of $D^0$ momentum to eliminate any biases due to
differences in the $D^0$ momentum spectra between the data and the MC.
The preliminary results are 
$A(K^+ K^-) = 0.0005 \pm 0.0218 ({\rm stat}) \pm 0.0084 
({\rm syst})$ and $A(\pi^+ \pi^-) = 0.0195 \pm 0.0322 ({\rm stat}) \pm 0.0084 
({\rm syst})$.

We use many different variations of the fit shapes, both empirical and 
analytical, to assess the systematic uncertainties due to the fitting 
procedure (0.69\%).
We also consider biases due to the detector material (0.07\%), the reconstruction
software (0.48\%), and forward--backward acceptance variations ($c \bar{c}$ 
pairs are not produced symmetrically in the forward/backward directions
in $e^+ e^-$ collisions at $\sqrt{s} \sim 10.6$ GeV, and the collision
point was not centered exactly in the middle of the detector) (0.014\%).

\begin{figure}
\centerline{
 \psfig{file=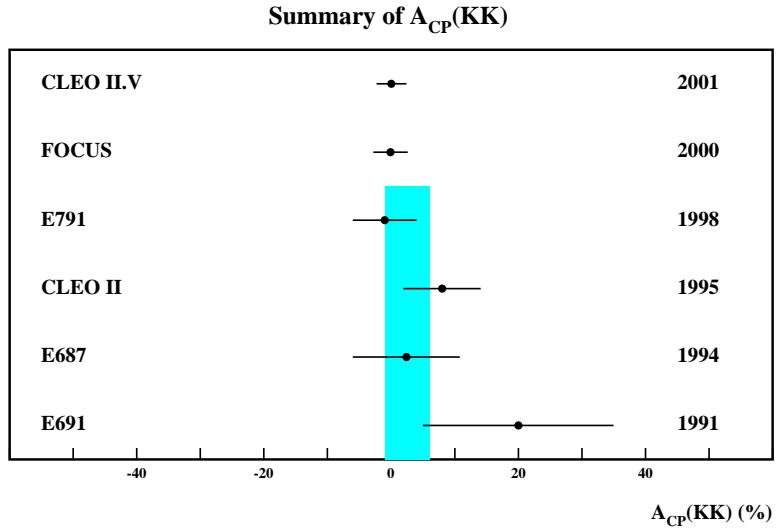,width=4.0in}}
\vspace{0.5cm}
\caption{\label{fig:hillkk} $A(K^+ K^-)$ from this analysis compared to previous
results.}
\end{figure}

\begin{figure}
\centerline{
 \psfig{file=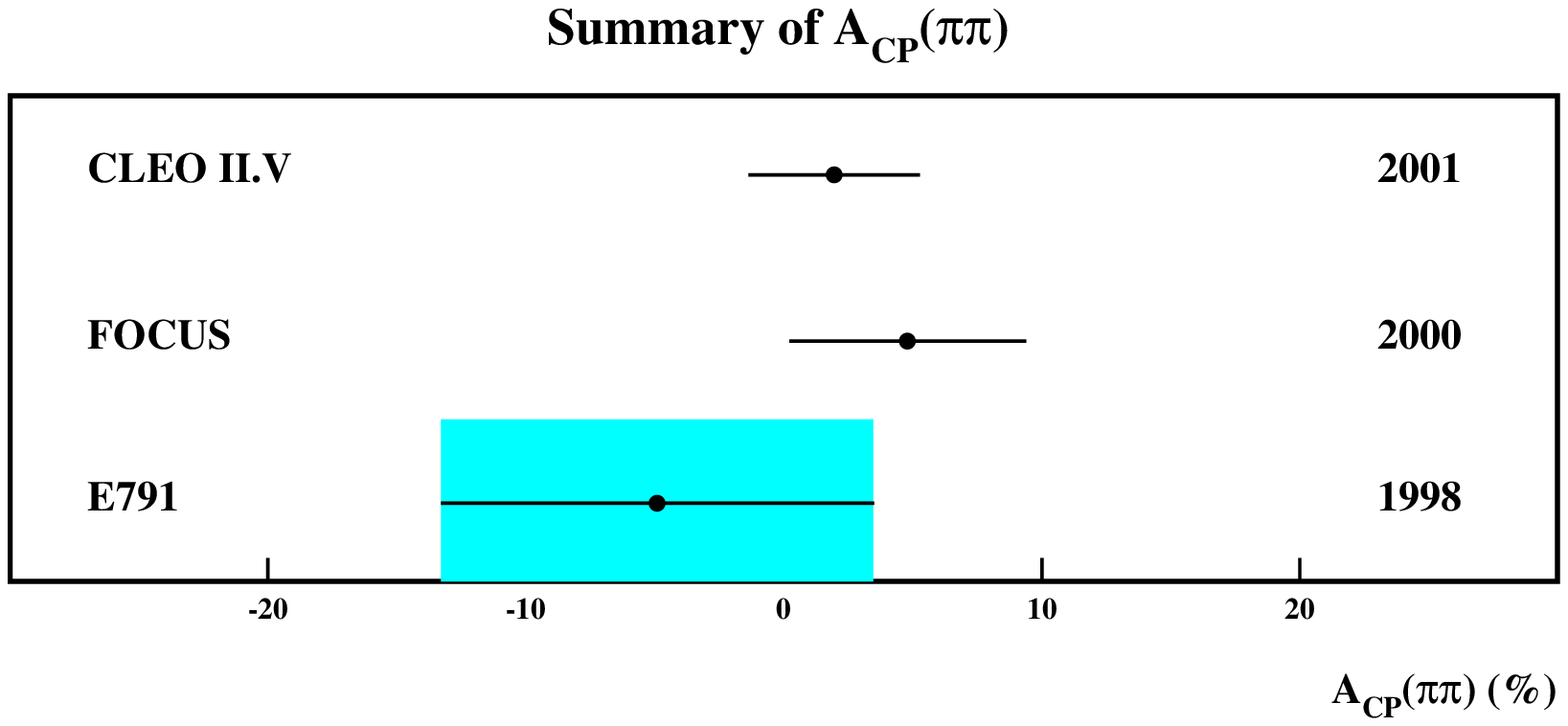,width=4.0in}}
\vspace{0.5cm}
\caption{\label{fig:hillpp} $A(\pi^+ \pi^-)$ from this analysis compared to 
previous results.}
\end{figure}

The measured asymmetries are consistent with zero, and no {\em CP} violation
is seen.  These results compared to previous measurements~\cite{kspi0,OLDCP} 
are shown in Figures~\ref{fig:hillkk} and~\ref{fig:hillpp}.
%
%

\subsection{Search for {\em CP} Violation in $D^0 \rightarrow K^0_{\rm
S} \pi^0$, $D^0 \rightarrow \pi^0 \pi^0$ and $D^0 \rightarrow K^0_{\rm S} 
K^0_{\rm S}$ decay}

This analysis~\cite{jaffe} differs from the other analyses 
presented in this paper in 
some of its reconstruction techniques and in the data set used.  
The $\pi^0 \pi^0$ and
$K^0_{\rm S} \pi^0$ final states do not provide sufficiently precise
directional information about their parent $D^0$ to use the intersection of the $D^0$ 
projection and the CESR luminous region to refit the
slow pion as described in the general experimental technique section. 
The $K^0_{\rm S} K^0_{\rm S}$ final state is treated the same for consistency.
This analysis uses the data from both the CLEO II and CLEO II.V
configurations
of the detector, totaling 13.7 ${\rm fb}^{-1}$ of $e^+ e^-$ collision
data at $\sqrt{s} \sim 10.6$ GeV.

\begin{figure}
\centerline{\psfig{file=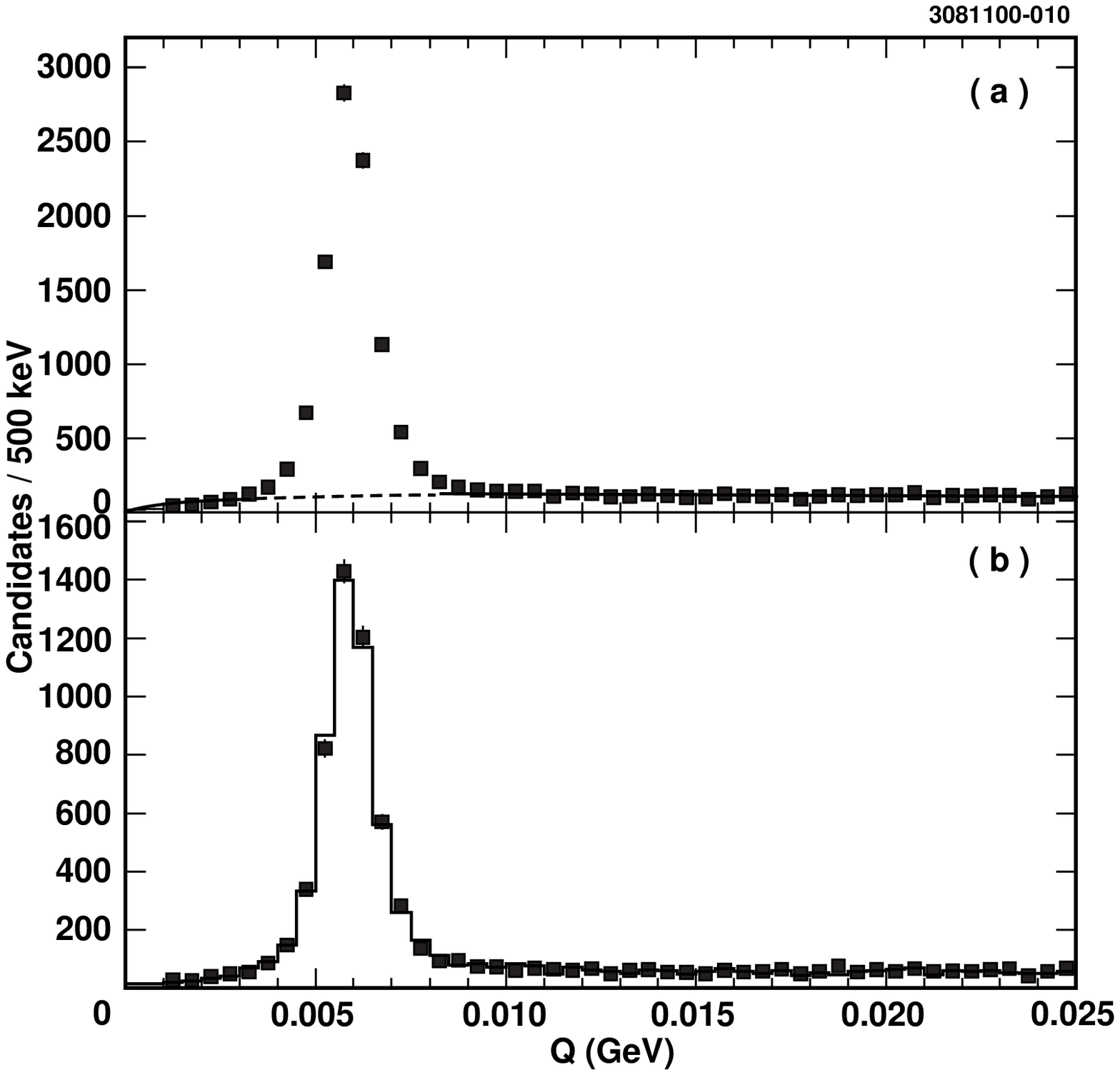,width=3.5in}}
\caption{\label{fig:k0pi0} (a) Fitted $Q$ distribution for 
$D^0 \rightarrow K_S^0 \pi^0$ candidates.  The points with error bars are 
the data, the solid line represents the background
and the dashed line shows the interpolation into the $Q$ signal region. 
(b) The Q distributions for $D^0 \rightarrow K_S^0 \pi^0$ (points) 
and $\overline{D^0} \rightarrow K_S^0 \pi^0$ (histogram) candidates.}
\end{figure}

\begin{figure}
\centerline{\psfig{file=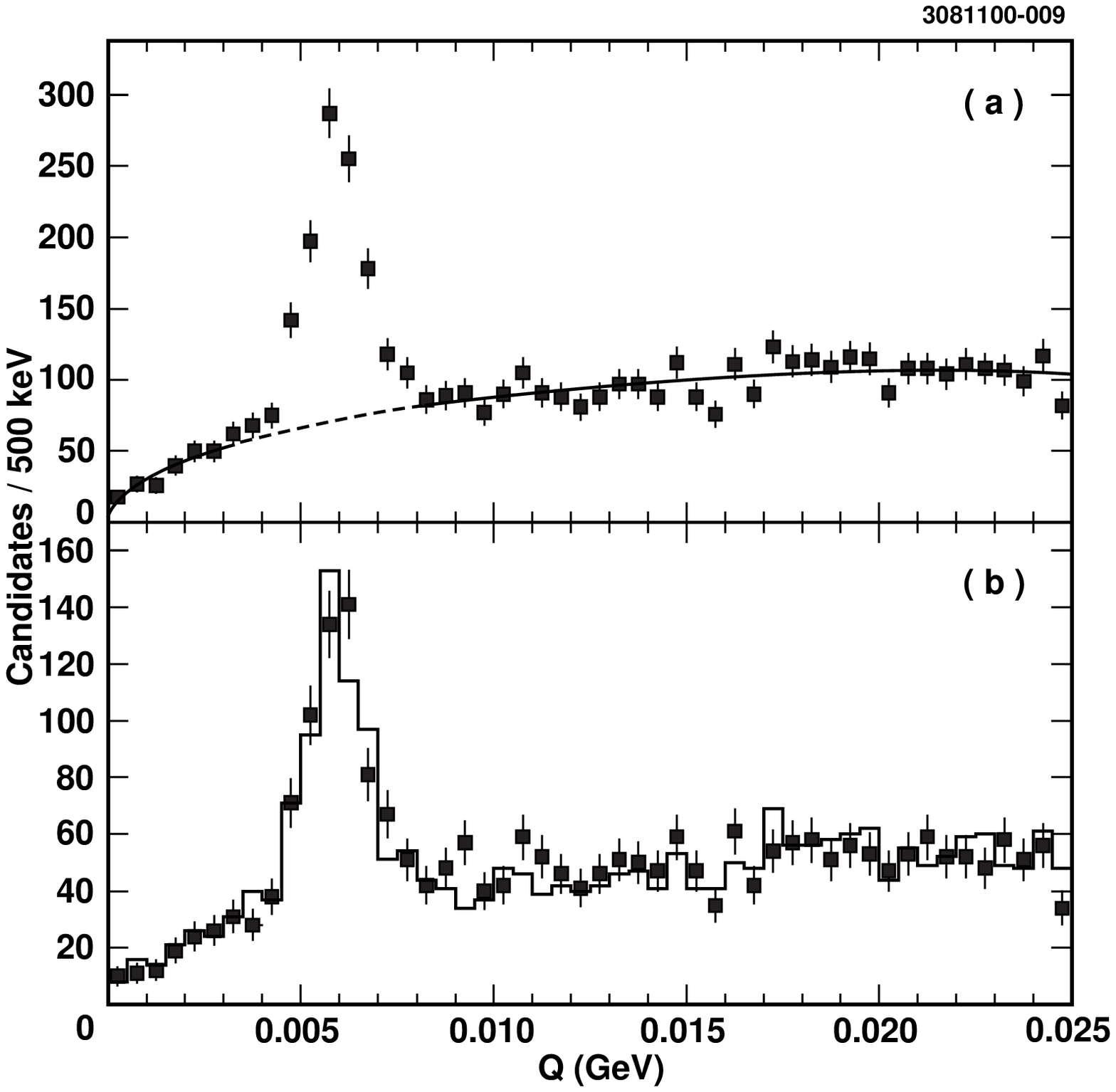,width=3.5in}}
\caption{\label{fig:pi0pi0} (a) Fitted $Q$ distribution for $D^0 \rightarrow \pi^0 \pi^0$
candidates.  The points with error bars are the data, the solid line represents the background
and the dashed line shows the interpolation into the $Q$ signal region. (b) The Q distributions for 
$D^0 \rightarrow \pi^0 \pi^0$ (points) and $\overline{D^0} \rightarrow \pi^0 \pi^0$ (histogram) 
candidates.}
\end{figure}

\begin{figure}
\centerline{\psfig{file=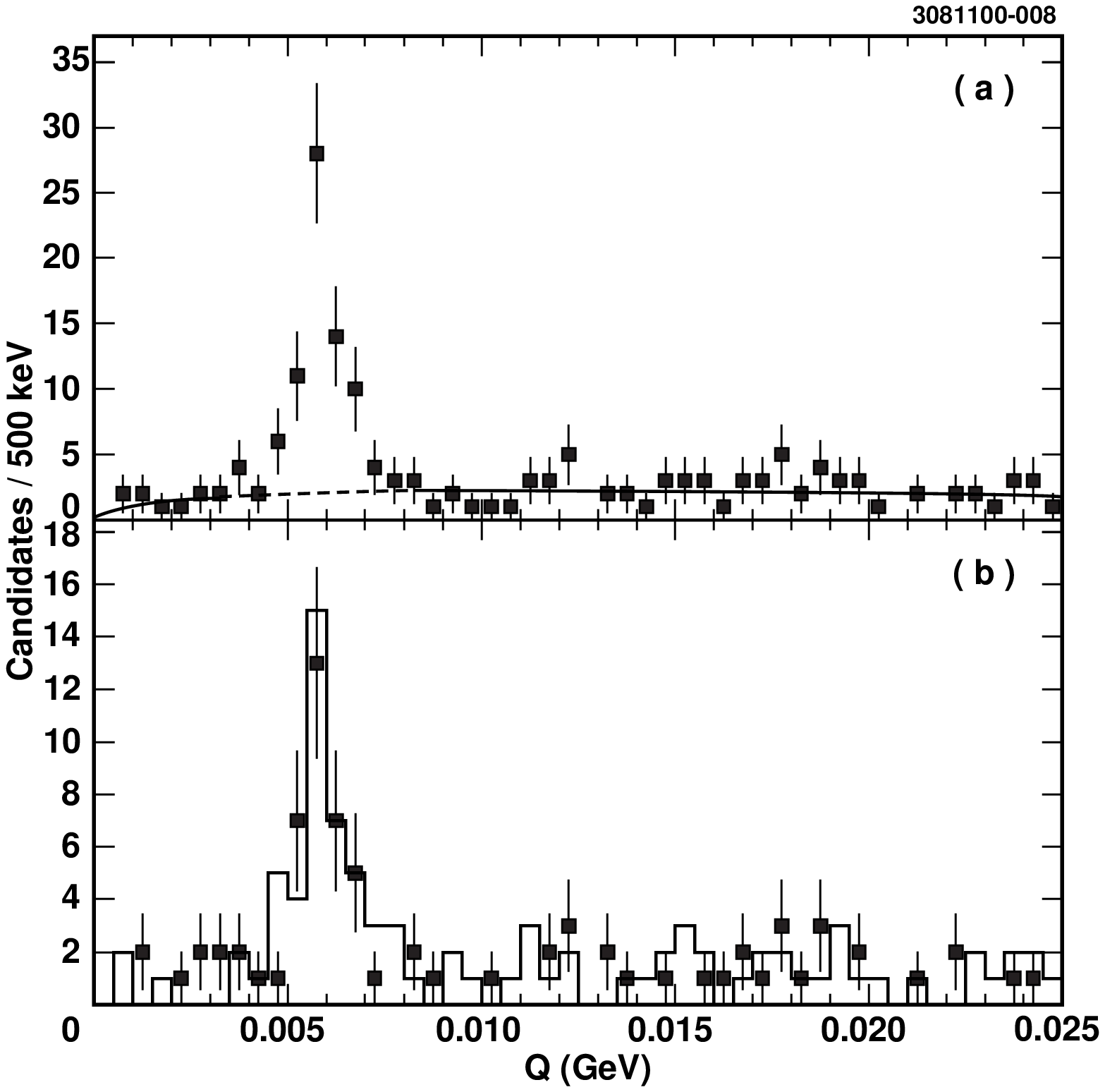,width=3.5in}}
\caption{\label{fig:k0k0} (a) Fitted $Q$ distribution for $D^0 \rightarrow K_S^0 K_S^0$
candidates.  The points with error bars are the data, the solid line represents the background
and the dashed line shows the interpolation into the $Q$ signal region. (b) The Q distributions for 
$D^0 \rightarrow K_S^0 K_S^0$ (points) and $\overline{D^0} \rightarrow K_S^0 K_S^0$ (histogram) 
candidates.}
\end{figure}

The $K^0_{\rm S}$ and $\pi^0$ candidates are constructed using only good 
quality tracks and showers.  The tracks (showers) whose combined 
invariant mass is close to the $K^0_{\rm S}$ ($\pi^0$) mass are
kinematically constrained to the $K^0_{\rm S}$ ($\pi^0$) mass, improving the 
$D^0$ mass resolution.  The tracks used to form $K^0_{\rm S}$
candidates are required to satisfy criteria designed to reduce
background from $D^0 \rightarrow \pi^+ \pi^- X$ decays and combinatorics.  
Candidate events with
reconstructed $D^0$ masses close to the known $D^0$ mass are selected to
determine the asymmetry, $A(f) = \left[ \Gamma (D^0 \rightarrow f) - \Gamma
(\overline{D^0} \rightarrow f)\right] / \left[ \Gamma (D^0 \rightarrow f) + 
\Gamma(\overline{D^0} \rightarrow f)\right]$.  
The $Q$ distributions of the candidates in the three 
decay modes are shown in Figures \ref{fig:k0pi0}, \ref{fig:pi0pi0}, and 
\ref{fig:k0k0}.  
A prominent peak indicative of $D^{\star +} \rightarrow D^0 \pi^+_{\rm s}$ 
decay is observed in all three distributions.
%
%
The total number of $D^0$ and $\overline{D^0}$ candidates for a given final
state is determined as follows.  We fit the $Q$ distribution outside of 
the signal region and interpolate the fit under the signal peak to 
determine the background in the signal region.  We subtract the background
in the signal region from the total number of events there to determine
the total number of signal events.  The background shape is
approximated as a non-relativistic threshold function with first and
second order relativistic corrections $B(Q) = aQ^{1/2} + b Q^{3/2} + c
Q^{5/2}$.

After background subtraction, we obtain $9099 \pm 153$ $K^0_{\rm S}
\pi^0$ candidates, $810 \pm 89$ $\pi^0 \pi^0$ candidates, and $65 \pm 14$ 
$K^0_{\rm S} K^0_{\rm S}$ candidates.  

The difference in the number of $D^0$ and $\overline{D^0}$ to a given final 
state is determined by taking the difference of the number of events in
the signal region, and the asymmetry is obtained by dividing by the
number of candidates determined above.  This method of determining the
asymmetry implicitly assumes that the background is symmetric.  

We have searched for any sources of false asymmetries: from the 
$\pi^+_{\rm s}$ finding (0.19\%), from the fitting (0.5\%), 
and from the backgrounds (0.35\% $K^0_{\rm S} \pi^0$, 0\% $\pi^0 \pi^0$
and 12\% $K^0_{\rm S} K^0_{\rm S}$).
We find no significant biases, but apply the measured corrections and
add their uncertainties to the total.  We obtain the results 
$A(K^0_{\rm S} \pi^0) = (+0.1 \pm 1.3)\%$, 
$A(\pi^0 \pi^0) = (+0.1 \pm 4.8)\%$ and 
$A(K^0_{\rm S} K^0_{\rm S}) = (-23 \pm 19)\%$ 
where the uncertainties contain the 
combined statistical and systematic uncertainties.  All systematic
uncertainties, except for the 0.5\% uncertainty assigned for possible 
bias in the fitting method, are determined from data and would be
reduced in future higher luminosity samples.  

All measured asymmetries are consistent with zero and no indication of
significant {\em CP} violation is observed.  This measurement of 
$A(K^0_{\rm S} \pi^0)$ is a significant improvement over previous results, 
and the other two asymmetries reported are first measurements.

\section{First Observation of Wrong-Sign $D^0 \rightarrow K^+ \pi^- \pi^0$ Decay}

$D^{0} \rightarrow K\pi\pi^{0}$ candidates are reconstructed using the
selection criteria described in Section~\ref{genmethod}, with additional 
requirements specific to
this analysis.  In particular, $\pi^{0}$ candidates with momenta greater than
340~M$e$V/$c$ are reconstructed from pairs of photons detected in the
CsI crystal calorimeter.  Backgrounds are reduced by requiring
specific ionization of the pion and kaon candidates to be consistent
with their respective hypotheses.

The RS mode was recently studied by CLEO\cite{bib:bergfeld} and  found
to have a rich Dalitz structure consisting  
of $\rho(770)^{+}$, $K^{*}(892)^{-}$, $\overline{K^{*}}(892)^{0}$, 
$\rho(1700)^{+}$, $\overline{K_{0}}(1430)^{0}$, $K_{0}(1430)^{-}$,
and $K^{*}(1680)^{-}$ resonances and non-resonant contributions.
Recent theoretical predictions based on 
U-spin symmetry arguments\cite{bib:GronauRosner} suggest that the wrong sign (WS)
channel will have a different resonant substructure than the right sign (RS).  
This analysis allows for different average WS and RS efficiencies 
resulting from different WS and RS kinematic distributions in the
calculation of the WS rate:
\begin{equation}
  \label{eq:rws}
  R_{WS} = \frac{\overline{\varepsilon}_{RS}}{\overline{\varepsilon}_{WS}}\cdot
  \frac{N_{WS}}{N_{RS}} .  
\end{equation}

\begin{figure}
\centerline{
 \psfig{file=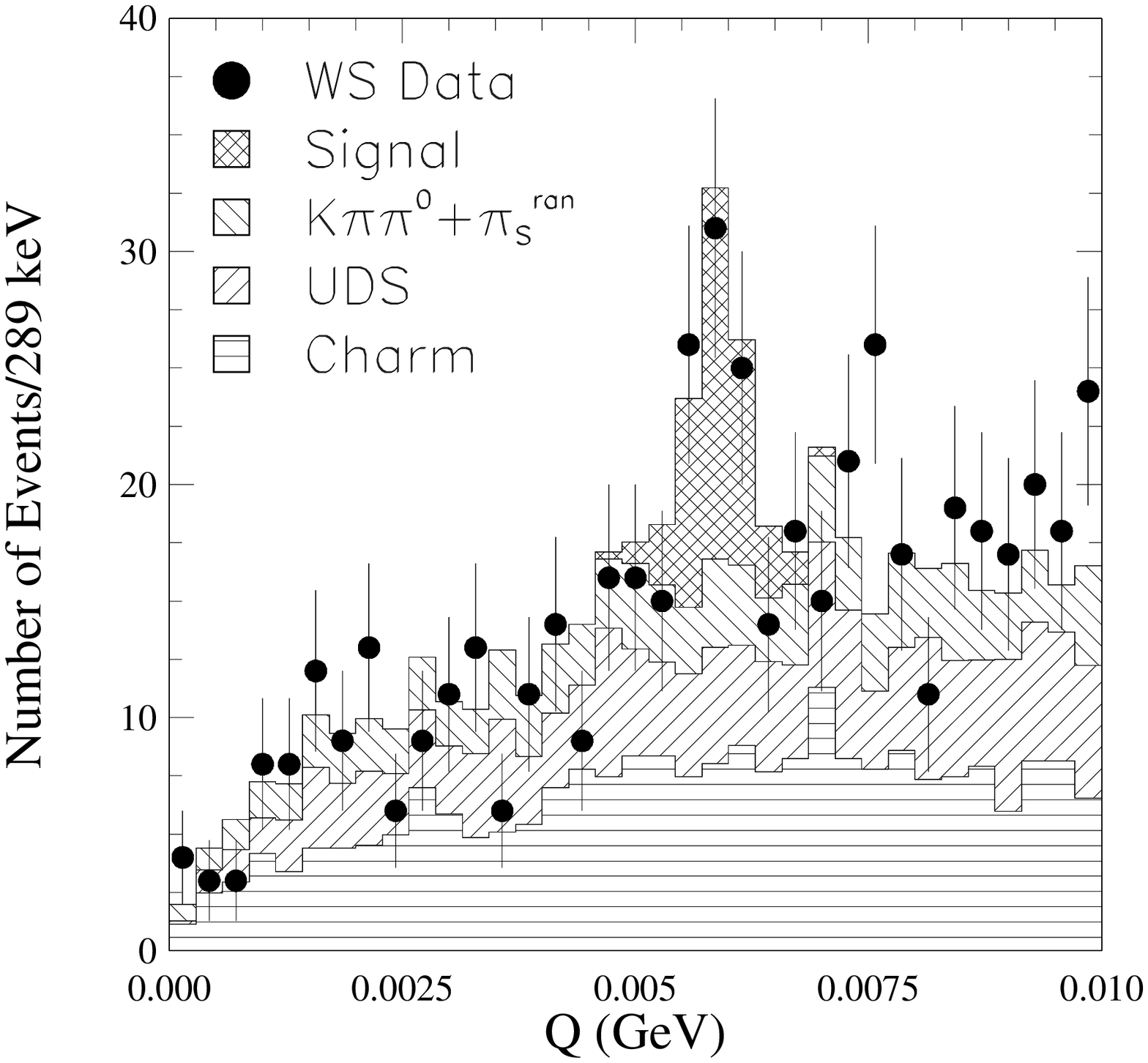,width=3.0in}
 \psfig{file=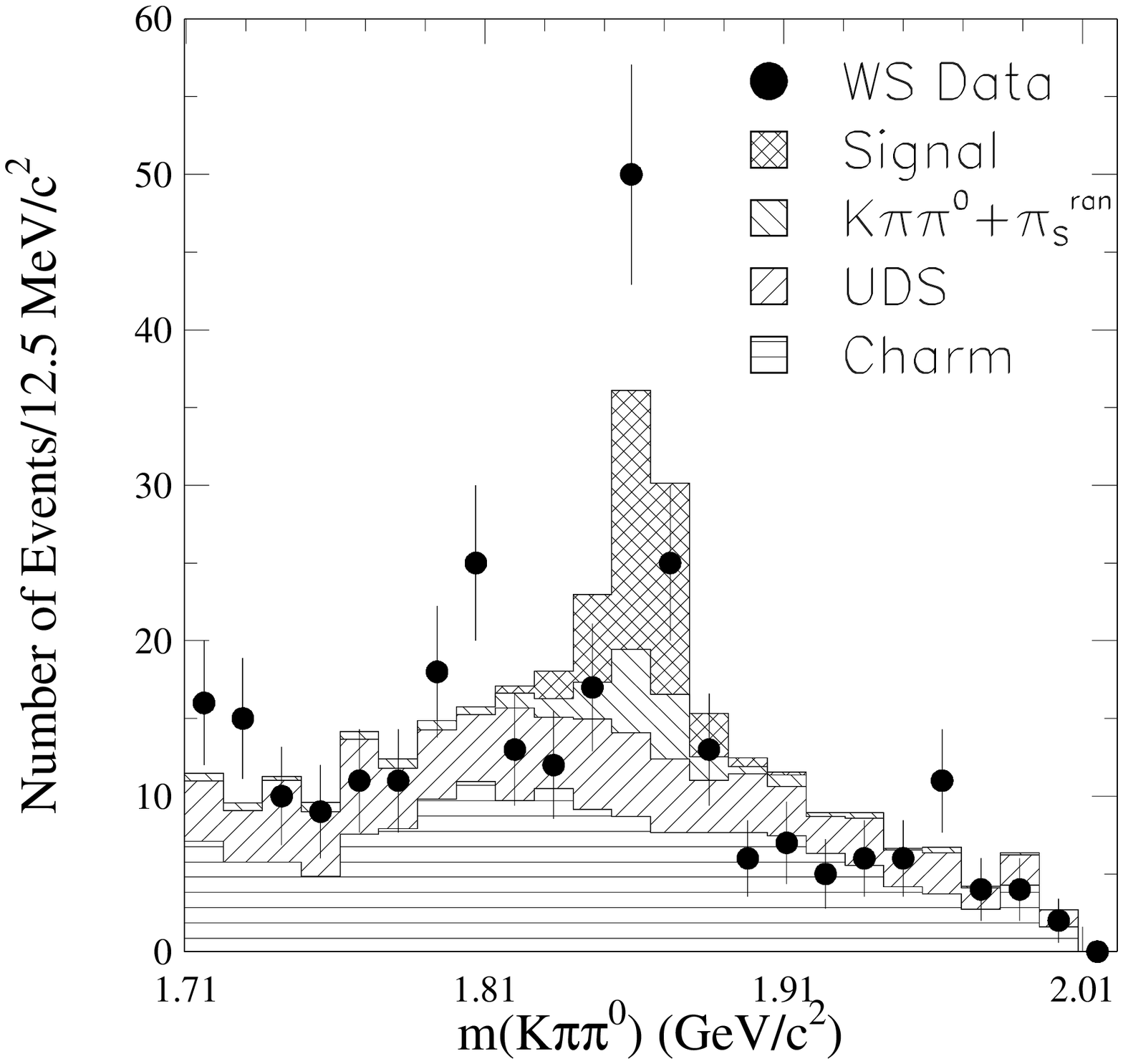,width=3.0in}
}
\caption{Results of the fit to the data to determine $N_{WS}/N_{RS}$.
  Projections in the variables a) $Q$ and b) $m(K\pi\pi^{0})$, after 
selecting the signal region (within two standard deviations) in the other
variable.}
\label{fig:alexmass}
\end{figure}

The ratio of yields in Eq.~(\ref{eq:rws}) is measured by performing a
maximum likelihood fit to the two-dimensional distribution in
$m(K\pi\pi)$ and $Q$.  The signal distribution in these variables is
taken from the RS data.  The backgrounds are broken down into three
categories: 1) RS $\overline{D^{0}}\rightarrow K\pi\pi^{0}$ decay
combined with an uncorrelated $\pi_{S}$, 2) combinations from
$e^{+}e^{-}\rightarrow u\overline{u}$, $d\overline{d}$, and
$s\overline{s}$, and 3) combinations from charm particle
decays other than the RS or WS signal modes.  The
background distributions are determined using a large Monte Carlo sample,
which corresponds to approximately eight times the integrated
luminosity of the data sample.  The $Q$--$m(K\pi\pi^{0})$ fit yields a WS signal of 
$38 \pm 9$ events and a ratio $N_{WS}/N_{RS}=0.0043^{+11}_{-10}$.
Projections of the data and fit results in slices through the signal region in
each variable are shown in Fig.~\ref{fig:alexmass}.  
The statistical significance of this signal is found to be 4.9
standard deviations. 

The average efficiency ratio in Eq.~(\ref{eq:rws}) is determined
using a fit to the Dalitz plot variables $m^{2}(K^{+}\pi^{-})$ and
$m^{2}(K^{+}\pi^{0})$ in the WS data.  In this fit, the amplitudes and
phases are initialized to the RS values, and those
corresponding to the $K^{*}(892)^{+}$ and $K^{*}(892)^{0}$ resonances 
are floated relative to the dominant $\rho(770)^{-}$ and other minor
contributions.  Combining the square of the
fitted amplitude function with a parameterization of the efficiency
determined using a large non-resonant Monte Carlo sample, we measure
an average efficiency ratio of  $1.00\pm 0.02 ({\rm stat})$.  Studies
are under way to examine the extent and the significance of this
surprising similarity between the RS and WS Dalitz plots.

The dominant systematic errors in this analysis come from the
uncertainty in the Monte Carlo background distributions (14\%),
uncertainty in the amplitudes and phases that are fixed in the fit (8\%), 
and uncertainty in the background Dalitz plot distributions (3\%).

Several powerful checks of the $Q$--$m(K\pi\pi^{0})$ and Dalitz plot
fits are performed in order to verify the validity of these results. 
Fits using specific background regions of the
$Q$--$m(K\pi\pi^{0})$ plane test the sensitivity to the
Monte Carlo background distributions.  The WS Dalitz plot is also fit
using hypotheses which include one of the $K^{*}(892)^{+}$,
$K^{*}(892)^{0}$, or $\rho(770)^{-}$ resonances, in order to provide an upper limit
on the error due to this fit.  Only the $K^{*}(892)^{0}$ hypothesis
leads to an efficiency ratio that differs significantly from one, but this
hypothesis is strongly disfavored by the data.

\begin{figure}
\centerline{\psfig{file=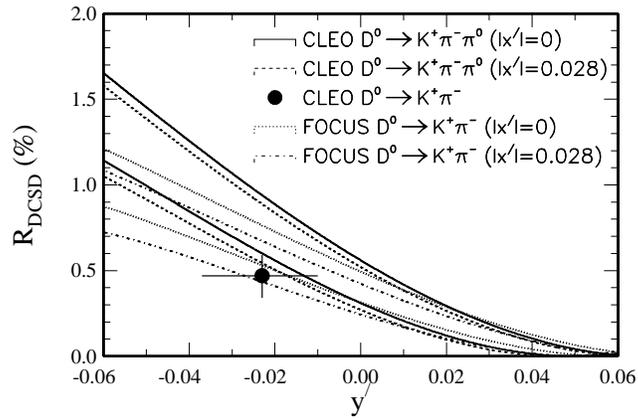,width=3.5in}}
\caption{Comparison of measured doubly-Cabibbo-suppressed rates as a 
function of $y^{\prime}$.  The variable $y^{\prime}$ may not be
the same for $D^{0}\rightarrow K^{+}\pi^{-}$ and 
$D^{0}\rightarrow K^{+}\pi^{-}\pi^{0}$.}
\label{fig:alexrws}
\end{figure}

We measure the wrong sign rate to be
\begin{equation}
  \label{eq:rwsresult}
     R_{WS} = 0.0043^{+0.0011}_{-0.0010}\ ({\rm stat}) \pm 0.0007\ ({\rm syst}) ~({\rm preliminary}).
\end{equation}
This result is consistent with the CLEO II.V\cite{kpi} and FOCUS 
\cite{FOCUS} $D^{0}\rightarrow K\pi$ measurements. 
This measurement of $R_{WS}$ can be used to obtain limits on
$R_{DCSD}$ as a  function of $y^\prime$ within the limits on
$x^{\prime}$ set by CLEO, as shown in Figure~\ref{fig:alexrws}.  
Work is in progress to use the lifetime distribution
of this sample to yield independent limits on $x^{\prime}$,
$y^{\prime}$, and $R_{DCSD}$.
%
%

\section{Search for $CP$ dependent lifetime differences due to 
$D^0-\overline{D^0}$ Mixing}

In the limit of no $CP$ violation in the neutral $D$ system
we can write the time dependent rate for $D \to f$, where $f$ is a
CP eigenstate, as
\begin{equation}
R(t) \propto e^{-t \Gamma(1 - y_{CP} \eta_{CP})}
\end{equation}
where $\Gamma$ is the average $D$ width, $\eta_{CP}$ is the $CP$
eigenvalue for $f$, and
\begin{equation}
y_{CP} = \frac{\Delta \Gamma}{2 \Gamma}
\end{equation}
where $\Delta \Gamma$ is the width difference between the physical
eigenstates of the neutral $D$.  In the limit that {\em CP} is conserved
in charm decays, $y_{CP} = y$, where $y$ is the mixing parameter defined
in Section~\ref{sec:intro}.  With the $CP$ asymmetry for $KK$ and
$\pi\pi$ consistent with zero as discussed in Section~\ref{sec:kkpp}, 
this limit is well motivated and we will simply use $y$ for the rest
of this section.  We can then express $y$ as
\begin{equation}
y = \frac{\tau_{\overline{CP}}}{\tau_{CP+}} - 1
\end{equation}
where $\tau_{\overline{CP}}$ is the lifetime of a $CP$ neutral state, such
as $K\pi$, and $\tau_{CP+}$ is the lifetime of a $CP$ even state,
such as $KK$ and $\pi\pi$.  Thus to measure $y$ we simply
take the ratio of the lifetimes of $D^0 \to K \pi$ to $D^0 \to KK$ and
$\pi\pi$.  Since the final states are very similar, our backgrounds 
are small and cross-feed among the final states is negligible many
of the sources of uncertainty cancel in the ratio.  A similar analysis
has recently been published by FOCUS comparing the
$K\pi$ and $KK$ final states\cite{FOCUSKK}.

\begin{figure}
\centerline{
 \psfig{file=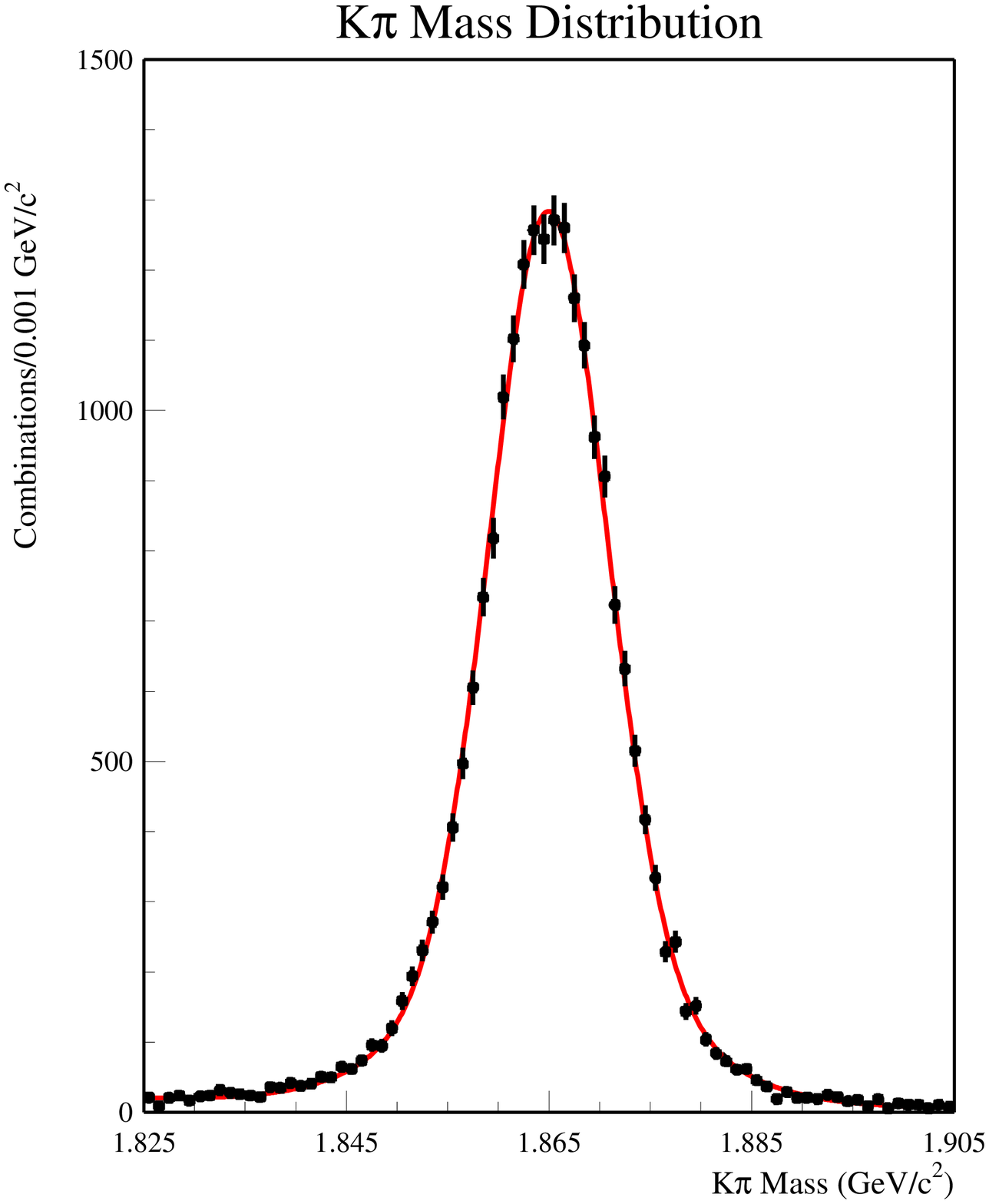,width=3.0in}
 \psfig{file=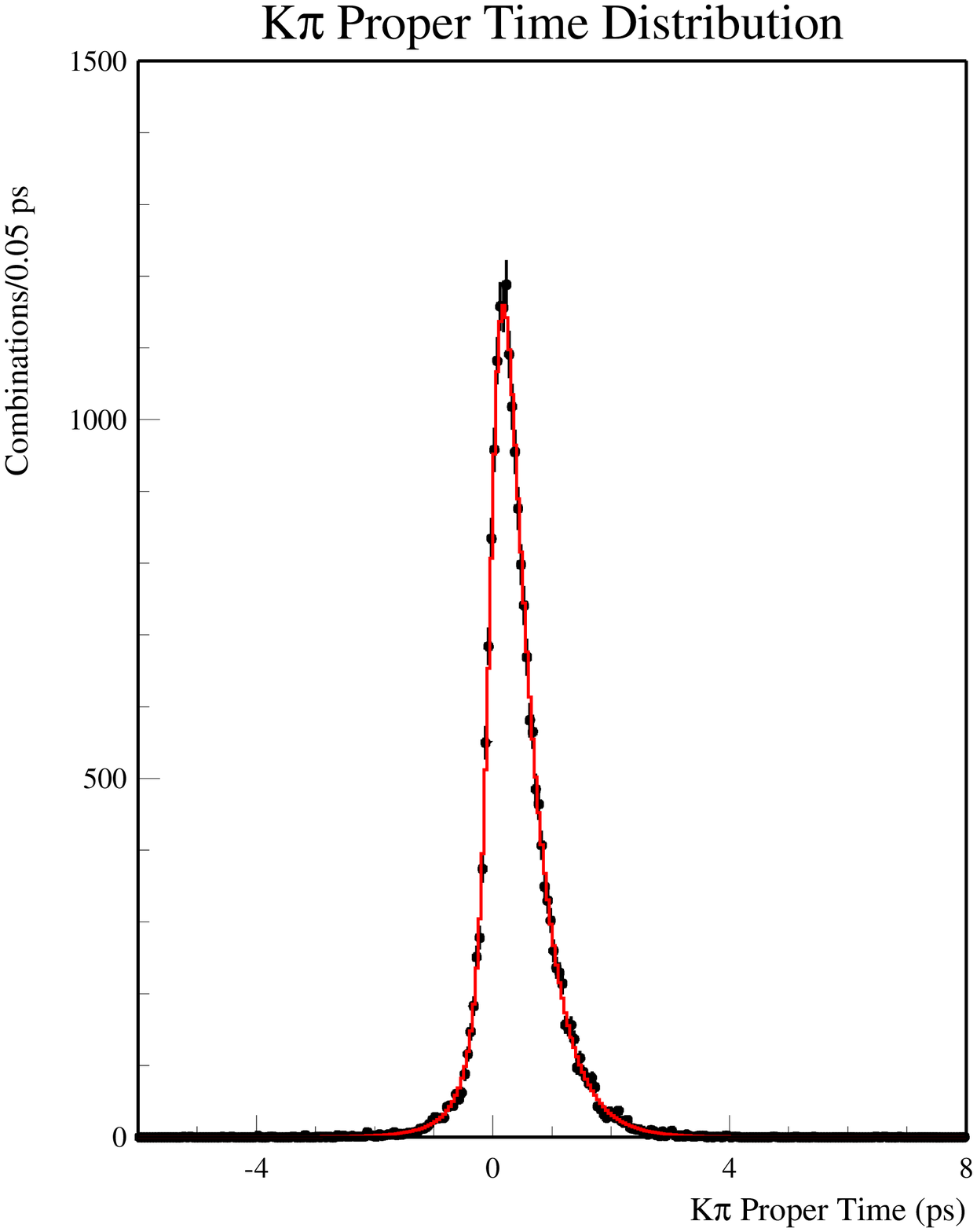,width=3.0in}
}
\caption{\label{fig:kpimass} The invariant mass (left) and proper time 
(right) distributions for $D^0\rightarrow K^- \pi^+$ candidates.
The curve is the fit described in the text.}
\end{figure}

\begin{figure}
\centerline{
 \psfig{file=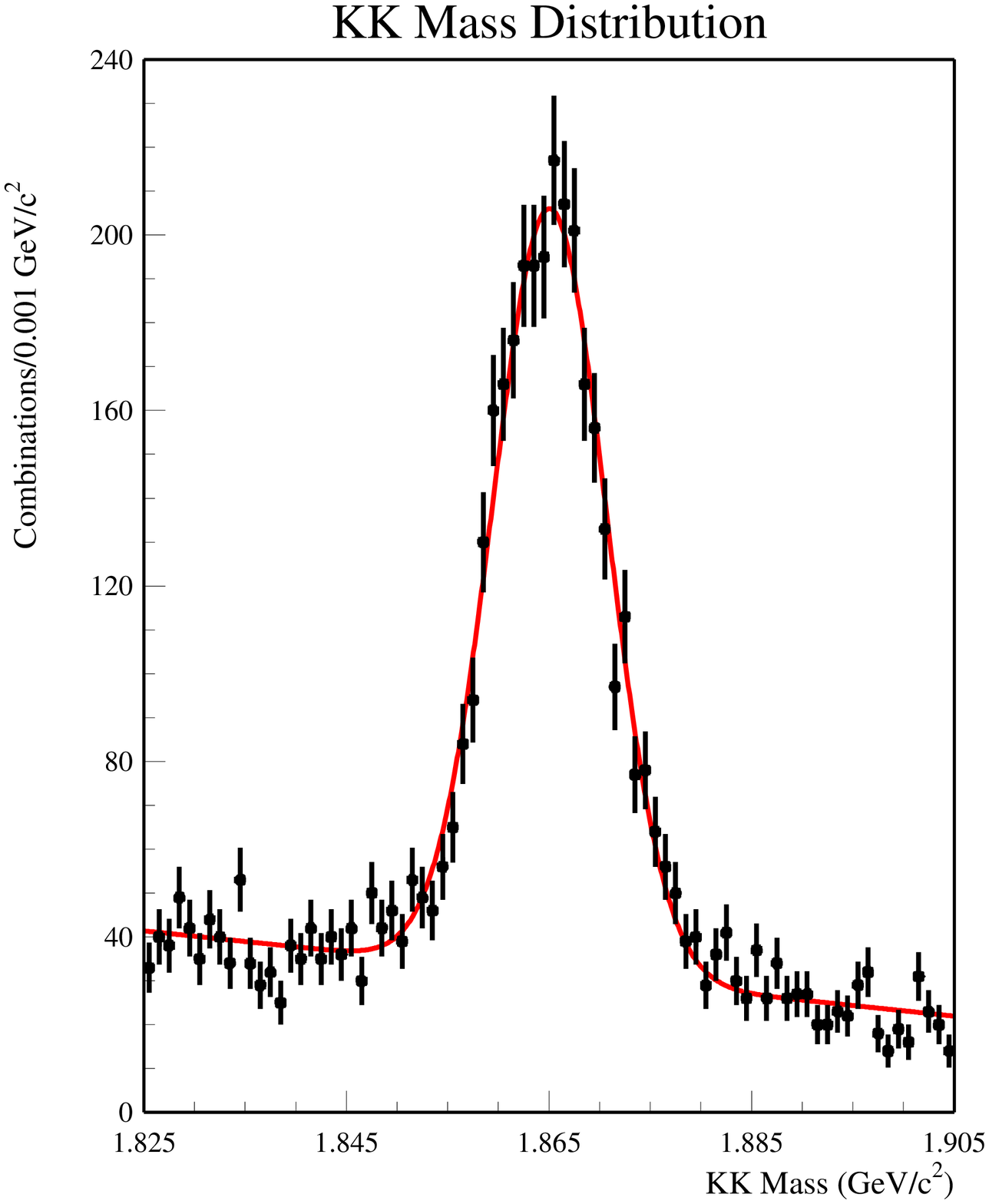,width=3.0in}
 \psfig{file=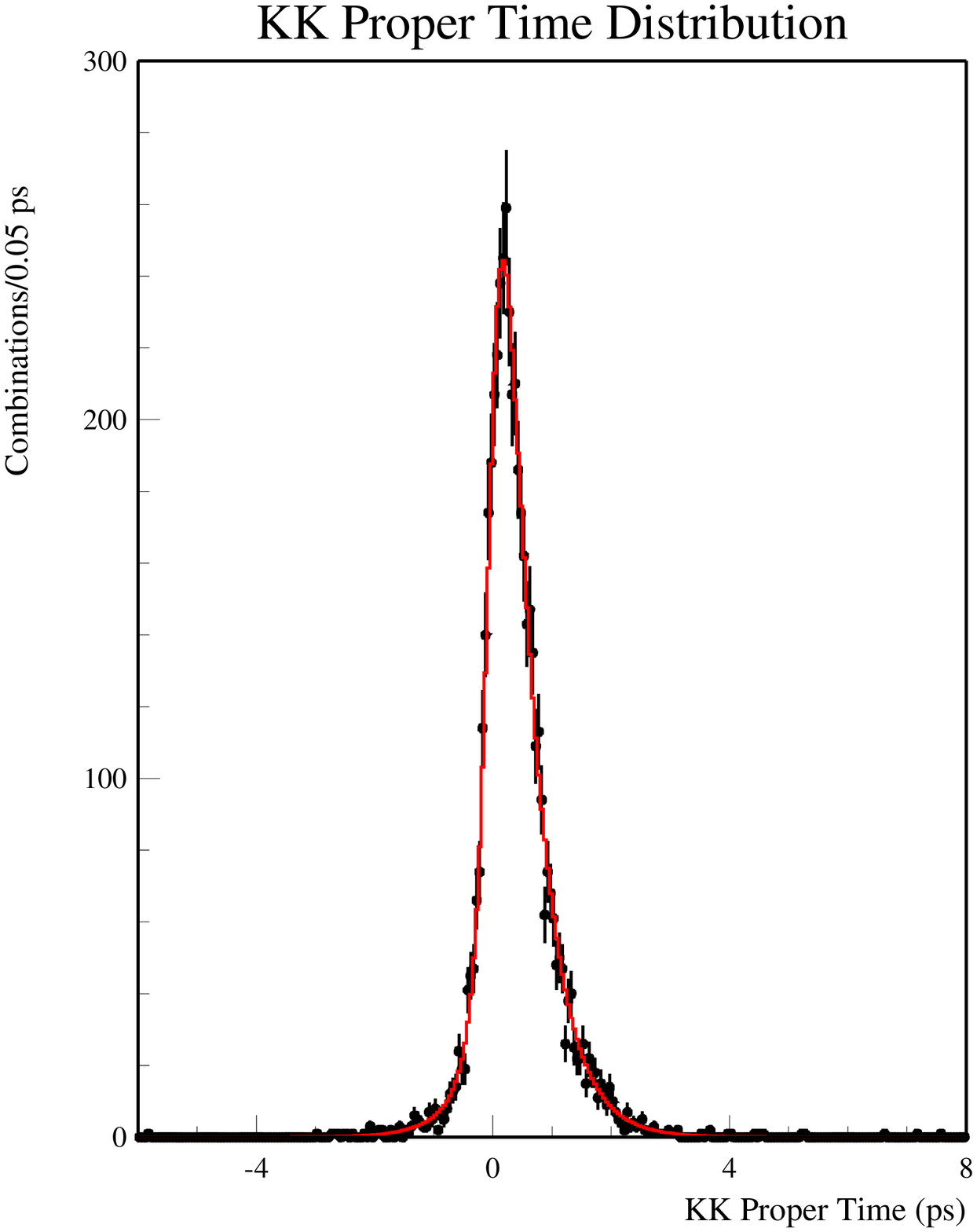,width=3.0in}
}
\caption{\label{fig:kkmass} The invariant mass (left) and proper time 
(right) distributions for $D^0\rightarrow K^+ K^-$ candidates.
The curve is the fit described in the text.}
\end{figure}

\begin{figure}
\centerline{
 \psfig{file=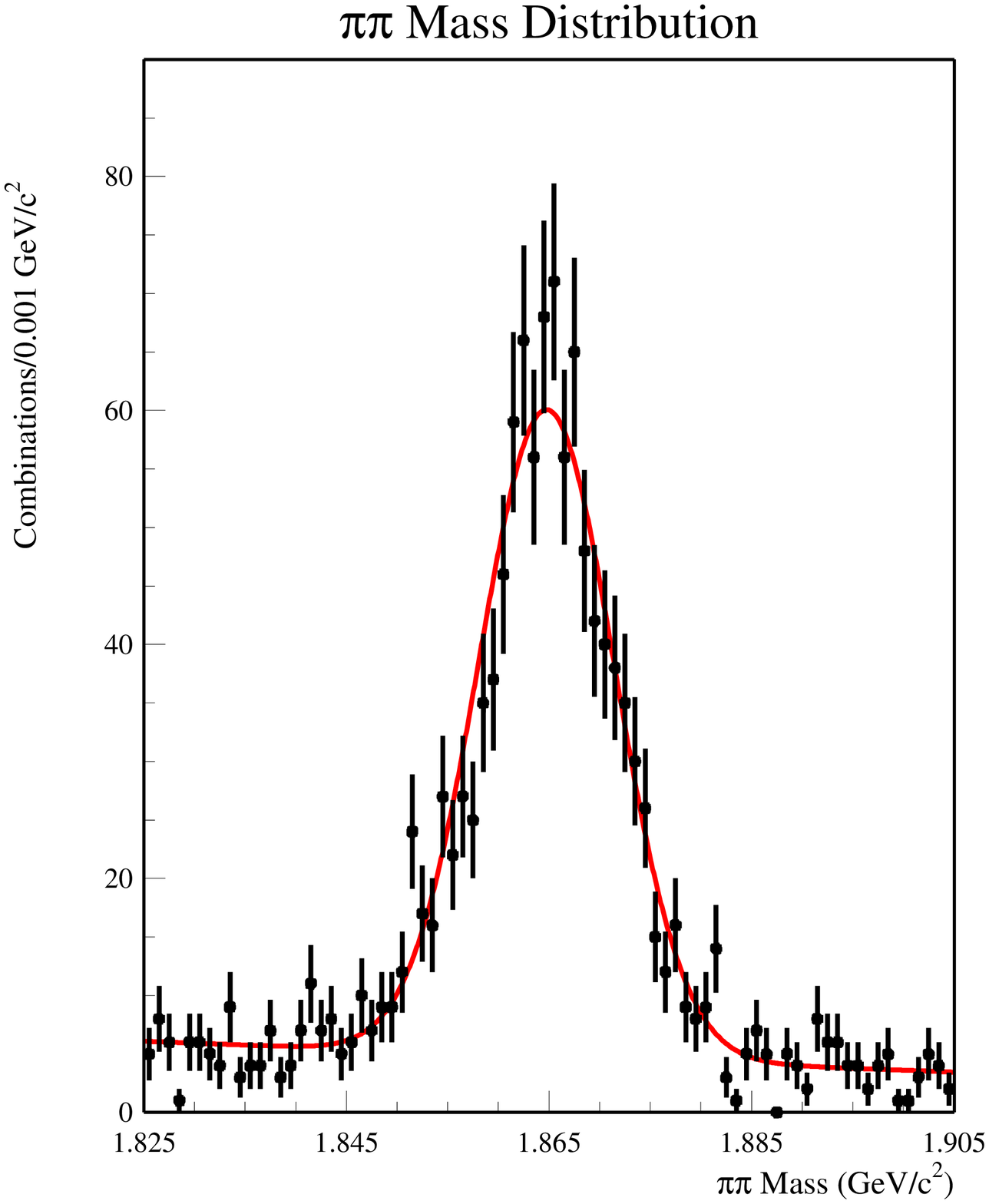,width=3.0in}
 \psfig{file=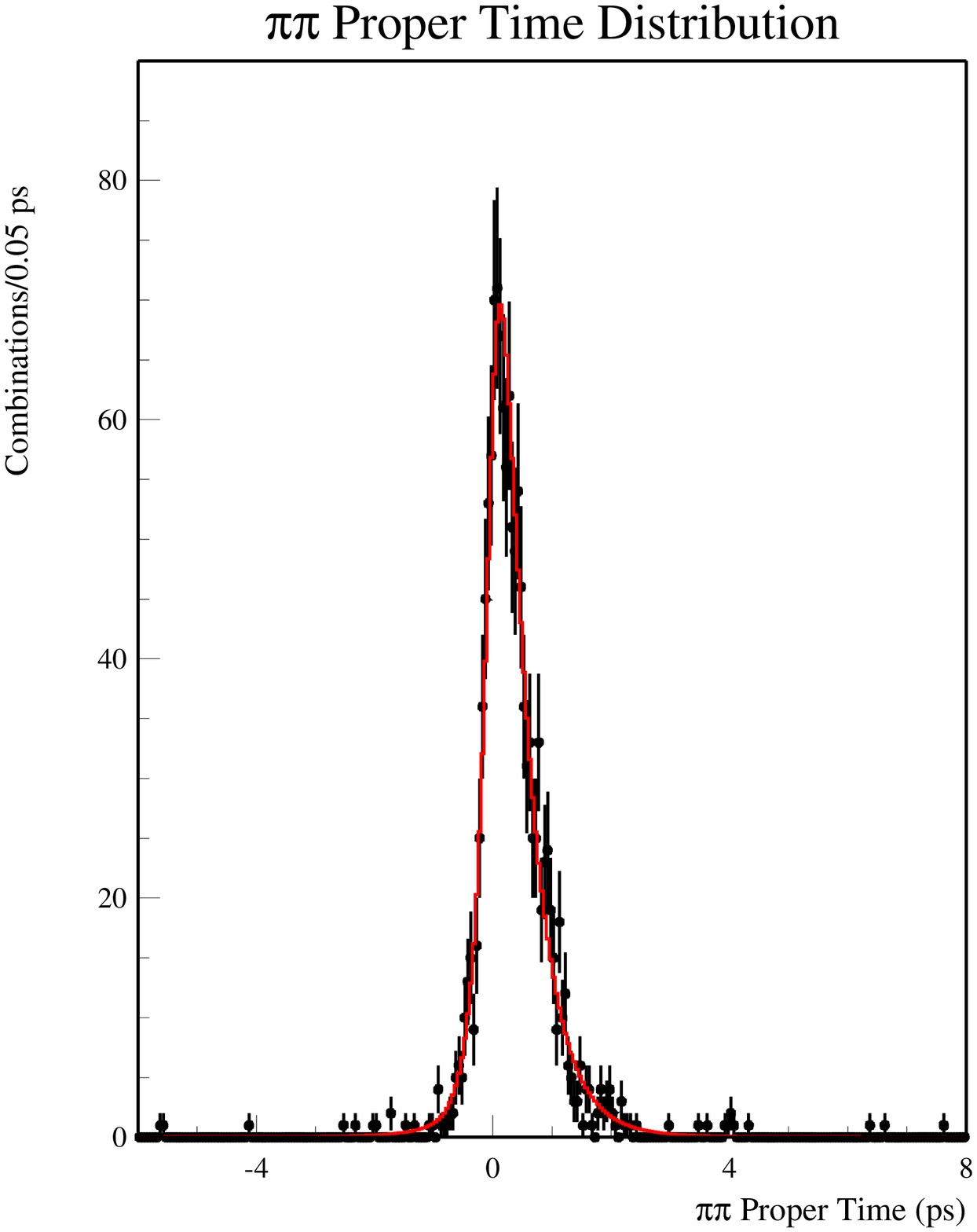,width=3.0in}
}
\caption{\label{fig:pipimass} The invariant mass (left) and proper time 
(right) distributions for $D^0\rightarrow \pi^+ \pi^-$ candidates.
The curve is the fit described in the text.}
\end{figure}

The left-most plots in Figures~\ref{fig:kpimass}, \ref{fig:kkmass},
and \ref{fig:pipimass} show the candidate $D^0$ masses for
the three samples selected as described in Section~\ref{genmethod}.  
Selecting events in a narrow region around
the $D$ mass, we fit their proper time distributions with an unbinned
maximum likelihood fit.  The proper time is measured as described in 
Section~\ref{genmethod}.
%
%
%
The signal likelihood function is 
an exponential convolved with three resolution Gaussians.  The 
width of the primary Gaussian is due to the propagation of
errors from the track fit to the flight distance and momentum
for each candidate.  The second and third Gaussians represent candidates
that have been mismeasured by the addition of spurious tracking hits or
due to hard non-Gaussian scatters in the material of the detector.
The first of these has its width determined in the fit to the copious 
$K\pi$ sample while the second is fixed to a large value, 8 ps.
The relative contribution of the two mismeasured signal resolutions
is determined using the $K\pi$ data sample and fixed for the 
$KK$ and $\pi\pi$ samples.
According to our simulation, the fraction of the second Gaussian
is about 4\% of the well measured signal and the third Gaussian
is less than 0.1\% of the signal.  The probability for a candidate to be signal
is determined by its measured mass, and is based on the fit to the 
mass distributions.

Background is considered to have contributions with both zero 
and non-zero lifetimes.
All parameters that describe the background are allowed to float in the fits 
except for the width of the widest Gaussian which is fixed to 8 ps.

The right-most plots in Figures~\ref{fig:kpimass}, \ref{fig:kkmass}, and
\ref{fig:pipimass} show the lifetime fits to the three
samples and the results are summarized in Table~\ref{tab:lifes}.
%
\begin{table}
\caption{Summary of the lifetime fits.  The parameters are those described in
the text.  Note that we have constrained the candidates to a $D^0$ mass of
1.86514 GeV, the Monte Carlo corrected weighted average of the data.  
The signal lifetime is highly dependent on the $D^0$ mass used
in the constraint, while the lifetime difference is not.  This technique yields
the smallest uncertainty in $y$, but it not optimal for measuring the absolute
$D^0$ lifetime and was not used in [34].}
\begin{center}
\begin{tabular}{|c|c|c|c|} \hline
Parameter           & $K\pi$              & $KK$              & $\pi\pi$ \\ \hline
Number of Signal    & $20272 \pm 178$     & $2463 \pm 65$     & $930 \pm 37$ \\
$\tau_{sig}$ (ps)   & $0.4046 \pm 0.0036$ & $0.411 \pm 0.012$ & $0.401 \pm 0.017$ \\
Background Frac (\%)& $8.8 \pm 0.2$       & $50.7 \pm 0.7$    & $29.1 \pm 1.3$ \\
Background Life Frac (\%) & $81.0 \pm 4.8$  & $85.7 \pm 2.9$  & $32.2 \pm 7.5$ \\ 
$\tau_{back}$ (ps)  & $0.376 \pm 0.030$   & $0.436 \pm 0.020$ & $0.56 \pm 0.15$\\ 
$f_{mis}$ \%        & $3.8 \pm 0.9$       & Fixed             & Fixed \\
$\sigma_{mis}$ (ps) & $0.590 \pm 0.079$   & Fixed             & Fixed \\ \hline
\end{tabular}
\end{center}
\label{tab:lifes}
\end{table}

Note that the background in all the samples has a large component
that has a lifetime consistent with the $D^0$ lifetime.  This agrees
with the prediction of our simulation that the background with 
lifetime is dominated by misreconstructed fragments of charm decays.

We calculate $y$ separately for the $KK$ and $\pi\pi$
samples.  Systematic uncertainties are dominated by the statistical uncertainty
in a Monte Carlo study used to determine small corrections, consistent with
zero, that are applied to the measured result to account for differences 
between measured and generated values of the lifetimes ($\pm 0.009$).  
Additional significant systematic uncertainties
come from variations in description of the background ($\pm 0.008$),
uncertainties in our model of the proper time resolution ($\pm 0.005$), and
details of the fit procedure ($\pm 0.005$), where the listed values are the
contribution to the average result.
Our preliminary results are 
\begin{equation}
y_{KK} = -0.019 \pm 0.029 ({\rm stat}) \pm 0.016 ({\rm syst})
\end{equation}
and 
\begin{equation}
y_{\pi\pi} = 0.005 \pm 0.043 ({\rm stat}) \pm 0.018 ({\rm syst}).
\end{equation}
We form a weighted average of the two to get
\begin{equation}
y =  -0.011 \pm 0.025 ({\rm stat}) \pm 0.014 ({\rm syst}) ~({\rm preliminary}).
\end{equation}

\begin{figure}
\centerline{
 \psfig{file=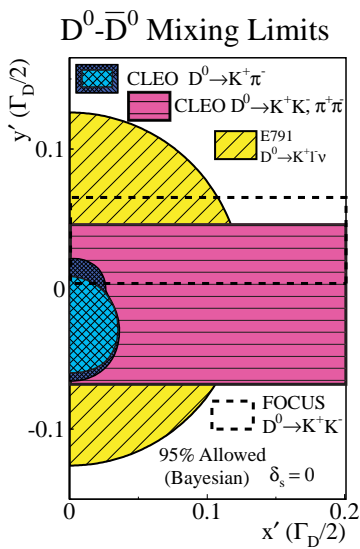,width=3.0in}}
\caption{\label{fig:newxy} Allowed regions, at 95\% CL, in the $x^\prime$ vs
$y^\prime$ plane for some recent results.
}
\end{figure}

This result is consistent with zero.  It is also consistent 
with our previous result from
wrong sign $K\pi$~\cite{kpi}, the FOCUS results from wrong sign 
$K\pi$~\cite{FOCUS}, and the
FOCUS results from $K\pi$ versus $KK$~\cite{FOCUSKK}. 
This result is shown along with some of the previous measurements in
Figure~\ref{fig:newxy}.
%

\section{Summary}
We  present preliminary results of several analyses 
searching for the effects of CP violation and mixing 
in the decay of $D^0$ mesons.  
We find no evidence of CP asymmetry in five different 
two-body decay modes of the $D^0$ to pairs of light pseudo-scalar mesons:
$A_{CP}(K^+ K^-) = (0.05 \pm 2.18 \pm 0.84)\%$, 
$A_{CP}(\pi^+ \pi^-) = (2.0 \pm 3.2 \pm 0.8)\%$,
$A_{CP}(K^0_{\rm S} \pi^0) = (+0.1 \pm 1.3)\%$, 
$A_{CP}(\pi^0 \pi^0) = (+0.1 \pm 4.8)\%$ and 
$A_{CP}(K^0_{\rm S} K^0_{\rm S}) = (-23 \pm 19)\%$.
We present the first measurement of the rate of wrong-sign 
$D^0 \rightarrow K^+ \pi^- \pi^0$ decay:
$R_{WS} = 0.0043^{+0.0011}_{-0.0010} \pm 0.0007$.
Finally, we describe a measurement of the mixing 
parameter $y={\Delta\Gamma\over 2 \Gamma}$ 
by searching for a lifetime difference between the CP 
neutral $K^+ \pi^-$ and the 
CP even $K^+K^-$ and $\pi^+\pi^-$ final states.  
Under the assumption that CP is conserved we find 
$y = -0.011 \pm 0.025 \pm 0.014$.

\section{Acknowledgments}
We gratefully acknowledge the effort of the CESR staff in providing us with
excellent luminosity and running conditions.
M. Selen thanks the PFF program of the NSF and the Research Corporation, 
and A.H. Mahmood thanks the Texas Advanced Research Program.
This work was supported by the National Science Foundation, the
U.S. Department of Energy, and the Natural Sciences and Engineering Research 
Council of Canada.


\begin{thebibliography}{99}
\bibitem{harry} H.N. Nelson, hep-ex/9908021. 
\bibitem{KTeV} KTeV Collaboration, A. Alavi-Harati {\it et al.}, Phys. Rev. Lett. 
	{\bf 83}, 22 (1999). 
\bibitem{NA48} NA48 Collaboration, V. Fanti {\it et al.}, Phys. Lett. B {\bf 465}, 
	335 (1999). 
\bibitem{BaBar} BaBar Collaboration, B. Aubert {\it et al.}, ``A study of time-
        dependent {\em CP}-asymmetries in $B^0_d \rightarrow J/\psi K^0_{\rm S}$
	and $B^0_d \rightarrow \psi(2S) K^0_{\rm S}$
        decays", BABAR-CONF-00/01, SLAC-PUB-8640, hep-ex/0008048. 
\bibitem{Belle} Belle Collaboration, H. Aihara, ``A measurement of {\em CP} 
	violation
        in $B^0_d$ meson decays with Belle", To be published in the proceedings
        of the 30th International Conference on High-Energy Physics
        (ICHEP 2000), Osaka, Japan, 27 Jul-2 Aug 2000, hep-ex/0010008. 
\bibitem{CDF} CDF Collaboration, F. Abe {\it et al.}, Phys. Rev. Lett. 
	{\bf 81}, 5513 (1998); T. Affolder {\it et al.} (CDF Collaboration), 
	Phys. Rev. D {\bf 61}, 072005 (2000). 
\bibitem{AGS} I.-H. Chiang {\it et al.}, AGS Experiment Proposal 926 (1996). 
\bibitem{KAMI} KAMI Collaboration, E. Cheu {\it et al.}, ``An expression of 
        interest to detect and measure the direct {\em CP} violating decay
        $K_L \rightarrow \pi^0 \nu \bar{\nu}$ and other rare decays at 
	Fermilab using the
        Main Injector", 22 September 1997, hep-ex/9709026. 
\bibitem{BTeV} ``Proposal for an Experiment to Measure Mixing, {\em CP} Violation
        and Rare Decays in Charm and Beauty Particle Decays at the 
        Fermilab Collider -BTeV", 15 may 2000, http://www-btev.fnal.gov/
        public\_documents/btev\_proposal/. 
\bibitem{LHCb} LHCb Collaboration, ``Technical Proposal A Large Hadron Collider
        Beauty Experiment for Precision Measurements of {\em CP} Violation and 
        Rare Decays", Printed at CERN, Geneva, Switzerland, ISBN 
        92-9083-123-5. 
\bibitem{Bucella} F. Bucella {\it et al.}, Phys. Rev. D {\bf 51}, 3478 (1995). 
\bibitem{Bigi} I.I. Bigi, ``Flavor dynamics - central mysteries of the
        Standard Model", To be published in the Proceedings of the 30th
        International Conference on High-Energy Physics (ICHEP 2000),
        Osaka, Japan, 27 Jul - 2 Aug 2000, hep-ex/0009021. 
\bibitem{Mixing} T.D. Lee, R. Oehme, and C.N. Yang, Phys. Rev. {\bf 106}, 340 (1957);
        A. Pais and S.B. Treiman, Phys. Rev. D {\bf 12}, 2744 (1975).  
	Our $x$ and $y$ are in
        terms of mixing amplitudes, while Pais and Treiman use the
        eigenvalues of the mixing Hamiltonian; both definitions agree in
        the limit of {\em CP} conservation.  In our convention, $y > 0$ 
	implies the real intermediate states with {\em CP} $= +1$, such as 
	$K^+ K^-$, make the larger contribution to $y$. 
\bibitem{GIM} S.L. Glashow, J. Illiopolous, and L. Maiani, Phys. Rev. D {\bf 2}, 
	1285 (1970). 
\bibitem{Leurer} M. Leurer, Y. Nir, and N. Seiberg, Nucl. Phys. {\bf B 420}, 468
        (1994); N. Arkani-Hamed {\it et al.}, hep-ph/9909326. 
\bibitem{kpi} CLEO Collaboration, R. Godang {\it et al.}, Phys. Rev. Lett. 
	{\bf 84}, 5038 (2000).
        $R_{WS} = (0.332^{+0.063}_{-0.065} \pm 0.040)\%$, 95\% CL 
	$(1/2){x^\prime}^2 < 0.041\%$
        and $-5.8\% < y < 1.0\%$. 
\bibitem{FOCUS} FOCUS Collaboration, J.M. Link {\it et al.}, hep-ex/0012048. 
\bibitem{PDG} C. Caso {\it et al.} (Particle Data Group), Eur. Phys. J. {\bf C 3}, 1 
	(1998). 
\bibitem{Treiman} S.B. Treiman and R.G. Sachs, Phys. Rev. {\bf 103}, 1545 (1956). 
\bibitem{Wolf} L. Wolfenstein, Phys. Rev. Lett. {\bf 75}, 2460 (1995); 
	T.E. Browder and S.
        Pakvasa, Phys. Lett. B {\bf 383}, 475 (1996); A.F. Falk, Y. Nir, and 
        A.A. Petrov, J. High Energy Phys. {\bf 9912}, 019 (1999). 
\bibitem{cleoii} Y. Kubota {\it et al.}, Nucl. Instrum. Methods Phys. Res. A
        {\bf 320}, 66 (1992). 
\bibitem{svx} T.S. Hill, Nucl. Instrum. Methods Phys. Res. A {\bf 418}, 32
        (1998). 
\bibitem{dr} D. Peterson, Nucl. Phys. {\bf B} (Proc. Suppl.) {\bf 54B}, 31 (1997). 
\bibitem{GEANT} R. Brun {\it et al.}, GEANT3 Users Guide, CERN DD/EE/84-1. 
\bibitem{charge} Charge conjugation is implied throughout, except where the 
	charge conjugate states are explicitly shown, such as in an asymmetry
	definition.
\bibitem{qwidth} This result is for the $D^0 \rightarrow K^+ \pi^-$ mode.  Other
	modes have similar widths since the uncertainty on the slow pion dominates
	the width of the $Q$ distribution.
\bibitem{GammaD*} First Measurement of $\Gamma(D^{\ast +})$, CLEO-CONF 01-02. 
\bibitem{cinabro} D. Cinabro {\it et al.}, Phys. Rev. E {\bf 57}, 1193 (1998).
\bibitem{kspi0} CLEO Collaboration, J. Bartelt {\it et al.}, Phys. Rev. D 
	{\bf 52}, 4860 (1995). 
\bibitem{OLDCP} FOCUS Collaboration, J.M Link {\it et al.}, Phys. Lett. B
	{\bf 491}, 232 (2000); Erratum-ibid. {\bf 495}, 443 (200); 
	E791 Collaboration, E.M.
	Aitala {\it et al.}, Phys. Lett. B {\bf 421}, 405
	(1998); E687 Collaboration, P.L. Frabetti {\it et al.}
	, Phys. Rev. D {\bf 50}, 2953 (1994); E691 Collaboration, 
	J.C. Anjos {\it et
	al.}, Phys. Rev. D {\bf 44}, 3371 (1991).
\bibitem{jaffe} CLEO Collaboration, G. Bonvicini {\it et al.}, to be published
	in Phys. Rev. D Rap. Comm.,
	CLNS 00/1708, hep-ex/0012054.
\bibitem{bib:bergfeld} CLEO Collaboration, S. Kopp {\it et al.}, hep-ex/0011065,
	submitted to Phys. Rev. D, November 2000.
\bibitem{bib:GronauRosner} M. Gronau, J. L. Rosner, hep-ph/0010237 (2000),
	submitted to Phys. Lett. B.
\bibitem{liftime} CLEO Collaboration, G. Bonvicini {\it et al.}, Phys. Rev. Lett.
	{\bf 82}, 4586 (1999).
\bibitem{FOCUSKK} FOCUS Collaboration, J.M. Link {\it et al.}, Phys. Lett. B
	{\bf 485}, 62 (2000).
\end{thebibliography}
\end{document}